\documentclass[prb,twocolumn]{revtex4}
\usepackage{graphicx}
\usepackage{dcolumn}
\usepackage{bm}
\usepackage{amsmath}
\usepackage{amsfonts}
\usepackage{psfrag}
\usepackage{graphicx}
\usepackage{color}

\newcommand{\Eq}[1]{Eq. (\ref{#1})}
\newcommand{\ma}[1]{a_{\mathbf{#1}}}
\newcommand{\mc}[1]{a^{\dagger}_{\mathbf{#1}}}
\newcommand{\ba}[1]{b_{\mathbf{#1}}}
\newcommand{\bc}[1]{b^{\dagger}_{\mathbf{#1}}}
\newcommand{\ca}[2]{c_{#1,\mathbf{#2}}}
\newcommand{\cc}[2]{c^{\dagger}_{#1, \mathbf{#2}}}

\begin{document}

\title{Terahertz lasing from intersubband polariton-polariton scattering in asymmetric quantum wells}

\author{Simone \surname{De Liberato}$^1$}
\author{Cristiano Ciuti$^2$}
\author{Chris C. Phillips$^3$}
\affiliation{$^1$ School of Physics and Astronomy, University of Southampton, Southampton, SO17 1BJ, United Kingdom}
\affiliation{$^2$ Laboratoire Mat\'eriaux et Ph\'enom\`enes Quantiques, Universit\'e Paris  Diderot-Paris 7 and CNRS, UMR 7162, 75013 Paris, France}
\affiliation{$^3$ Physics Department, Imperial College London, London SW7 2AZ, United Kingdom}

\begin{abstract}
Electric dipole transitions between different cavity polariton branches or between dressed atomic states with the same excitation number are strictly forbidden in centro-symmetric systems. 
For doped quantum wells in semiconductor microcavities, the strong coupling between an intersubband transition in the conduction band and a  cavity mode produces two branches of 
intersubband cavity polaritons, whose normal-mode energy splitting is tunable and can be in the terahertz region. 
Here, we show that, by using asymmetric quantum wells, it is possible to have allowed dipolar transitions between different polaritonic branches, leading to the emission of terahertz photons.   We present a quantum field theory for such a system and predict that high-efficiency, widely tunable terahertz lasing can be obtained.
\end{abstract}

\maketitle

The resonant coupling between a  photon mode and an electronic transition can produce hybrid light-matter eigenstates, such as the well-known dressed states in atomic cavity quantum electrodynamics  or the
polariton excitations in solid-state systems. 
In general, if one considers a doublet of dressed states, the electric dipole transition between these two states is strictly forbidden in any centro-symmetric system \cite{Savenko12}.  Analogously, in the case of cavity polaritons,  a transition from the upper to the lower polariton branch cannot be accompanied by any photon emission and can be only of non-radiative origin (e.g., via phonon emission \cite{DeLiberato09b}).  
Recently though, it has been shown that, in exciton-polariton systems,  the application of an electric field produces hybridisation of exciton states with different parity, thus breaking such a selection rule \cite{Kavokin10,Savenko11,delValle11,Kavokin12}. This enables transitions between different exciton-polariton states and in principle paves the way to integrated terahertz (THz) sources where the excitation is provided by a near-infrared pump with potentially significant quantum efficiencies (up to $1.5\%$). 

Indeed, sources emitting in the THz region of the electromagnetic spectrum are a subject of intense investigations both for fundamental physics and applications. 
In fact the THz region represents a technological gap that can be covered only with difficulty, either with electronic devices (at lower frequencies), or with lasers (at higher frequencies).
A great research effort has been accomplished to extend semiconductor quantum cascade lasers \cite{Faist94}
to the THz region \cite{Tredicucci02}: in such devices transitions occurs from quantized subbands in the conduction band of suitably designed multiple quantum well structures. 
Although THz quantum cascade lasers are developing \cite{Chassagneux09}, fundamental limitations occur because the radiative lifetimes of the excited electronic states are long compared to the intrinsic fast non-radiative recombination channels that arise from efficient electron-phonon scattering (at least for incoherent electronic excitations \cite{DeLiberato09a}).
In the case of intersubband transitions, it is possible to reach the strong light-matter coupling regime by embedding a structure containing multiple doped quantum wells in microcavity resonators  \cite{Dini03,Ciuti05,Anappara09,Todorov10,Pereira07}, leading
to the creation of so-called intersubband cavity polariton modes. In such a system,  it is possible to control in-situ the value of the polariton splitting \cite{Anappara05,Anappara07,Gunter09,Zanotto12}  and it is possible to engineer the intersubband transitions in a very flexible way \cite{Geiser12,Delteil12}.  A fundamental unexplored question is whether in such a class of intersubband systems it would be possible to have efficient radiative transitions between polariton branches.

In this paper, we show that breaking the wavefunction symmetry using asymmetric, doped quantum wells embedded into a planar photonic cavity, enables radiative transitions between different intersubband polariton branches, leading to intrinsically efficient, widely tunable THz lasing. We develop a quantum theory describing such  two-photon strongly coupled process and determine the overall efficiency of the THz emission, in a way that takes into account the nonbosonicity of intersubband excitations  \cite{Combescot07, Luc13, DeLiberato09b}. 
We have calculated, both analytically and numerically, the intrinsic quantum efficiency for the considered process showing that, for realistic parameters, unprecedented quantum efficiencies are achievable.

\begin{figure}[t!]
\begin{center}
\hspace{-1cm}
\includegraphics[width=8.0cm]{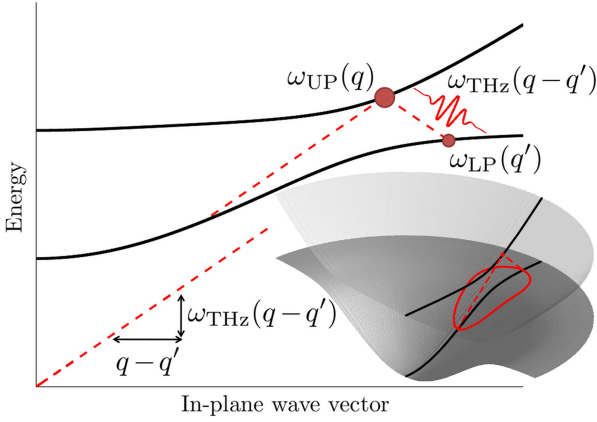}
\caption{\label{Sketch} Typical energy dispersion of the lower and upper intersubband polariton branches in a planar metallic microcavity (black solid lines), and of its TEM THz mode (red dashed line starting at the origin of the axis). The process described in the text, in which a pumped upper polariton scatters into a lower one by emitting a photon is shown. The red dots highlight the forward scattering channel, such that $\mathbf{q}$ and $\mathbf{q'}$ are parallel. The red dashed lines originating from the pumped upper polariton mode are a two dimensional cut of the lightcone available for spontaneous emission (in the three-dimensional image all the final states, given by the intersection of the emission lightcone with the lower polariton two dimensional manifold, are marked in red).  
}
\end{center}
\end{figure}
 
Let us start by considering the quantum Hamiltonian describing two parallel semiconductor conduction subbands coupled to a quasi-resonant microcavity photon mode, in the rotating wave approximation (RWA), namely
\begin{eqnarray}
\label{Hlm}
H&=&\sum_{\mu\in\{1,2\},\mathbf{k}}\hbar\omega_{\mu}(k)\cc{\mu}{k}\ca{\mu}{k}+\sum_{\mathbf{q}}\hbar\omega_{\text{ph}}(q)\mc{q}\ma{q}
\\&&+\sum_{\mathbf{k,q}} \hbar\chi(q)\mc{q}\cc{1}{k}\ca{2}{k+q}+ \hbar\chi(q)\ma{q}\cc{2}{k+q}\ca{1}{k},
\nonumber
\end{eqnarray}
where $\ca{\mu}{k}^\dagger$ and $a_{\mathbf{q}}^\dagger$  are, respectively, the creation operators for an electron in the quantum well conduction subband $\mu$ with in-plane wave vector $\mathbf{k}$ and a microcavity photon with in-plane wave vector $\mathbf{q}$.  Their respective energies are $\hbar\omega_{\mu}(k)$ and $\hbar\omega_{\text{ph}}(q)$,
with $\hbar\omega_{2}(k)=\hbar\omega_{1}(k)+\hbar\omega_{12}$. The coefficient $\hbar\chi(q)$ quantifies the light-matter coupling and it is proportional to the intersubband dipole $z_{12}$, where we define the matrix elements
\begin{eqnarray}
\label{zdef}
 z_{\mu\mu'}=\int dz\,z\, \psi_{\mu}(z)\bar{\psi}_{\mu'}(z), 
 \end{eqnarray}
 $\psi_{\mu}(z)e^{i\mathbf{\boldsymbol{{\rho}}\, k}}/\sqrt{S}$
is the wave function of an electron in subband $\mu$ with in-plane wave vector $\mathbf{k}$, $S$ is the sample surface and $({z},\boldsymbol{{\rho}})$ are the out- and in-plane components of the electron position.

Notwithstanding the fact that Hamiltonian in \Eq{Hlm} neglects Coulomb interaction, it has been the starting point for many quantum descriptions of intersubband polaritions, fitting experimental resonances with unprecedented precision (e.g., a relative RMS error $<1\%$ in Ref. [\onlinecite{Anappara09}]). This has been possible thanks to the fact that, as proved using various theoretical approaches \cite{Nikonov97,Todorov10,DeLiberato12,Kyriienko12,Luc13}, in the mid-infrared regime the Coulomb interaction only accounts for a negligible renormalisation of the intersubband energy $\hbar\omega_{12}$. 
All the interactions are spin conserving and are between states in the same quantum well, so for simplicity we omit the spin and quantum well indices. Due to the intersubband transition selection rules, only the Transverse Magnetic (TM) microcavity mode couples to electrons. Introducing the (approximately bosonic) bright intersubband operator $\bc{q}$  \cite{Ciuti05}, 
\begin{eqnarray}
\label{bc}
\bc{q}&=&\frac{1}{\sqrt{N}}\sum_{\mathbf{k}}\cc{2}{k+q}\ca{1}{k},
\end{eqnarray}
the  \Eq{Hlm} can be rewritten as  
\begin{eqnarray}
\label{HB}
H&=&\hbar\sum_{\mathbf{q}} \omega_{\text{ph}}(q)\mc{q}\ma{q}+\omega_{12}\bc{q}\ba{q}
+\Omega(q)(\mc{q}\ba{q}+ \bc{q}\ma{q}),\quad\,\,  \nonumber\\
&=& \sum_{j \in \{\text{LP},\text{UP} \},\mathbf{q}}  \hbar \omega_{j}(q) p^{\dagger}_{j,\mathbf{q}} p_{j,\mathbf{q}},
\end{eqnarray}
 where $N=n_{\text{QW}}N_{\text{2DEG}}S$ is the total number of electrons in the structure ($n_{QW}$ is the number of quantum wells and $N_{\text{2DEG}}$ is the density of the two-dimensional electron gas) and
$\Omega(q)=\sqrt{N}\chi(q)$ is the vacuum Rabi frequency.
The Hamiltonian in \Eq{HB} can be diagonalized in terms of the normal-mode polariton operators 
\begin{eqnarray}
\label{pc}
p_{j,\mathbf{q}}^\dagger={x}_{j,q} a_{\mathbf{q}}^\dagger+{y}_{j,q} b_{\mathbf{q}}^\dagger,
\end{eqnarray} 
where ${x}_{j,q}$ and ${y}_{j,q}$ 
are the (real) Hopfield coefficients that denote the light and matter fractions of the polaritonic excitations respectively.
 
Note that we can only use this simple diagonalisation, which leaves the creation and annihilation operators uncoupled, because we used the RWA in Eqs. (\ref{Hlm}) and (\ref{HB}). For practical applications, this approximation has been shown to be valid at least up to vacuum Rabi frequencies of $\simeq 0.1\omega_{12}$ \cite{Anappara09}.
In Fig. \ref{Sketch}, a sketch of the typical energy-momentum dispersion for the lower and upper polariton branches is shown. 

For intersubband cavity polaritons with energies in the mid-infrared,
transitions between the upper and lower polariton branch are typically in the THz region of the spectrum. 
The cavity also contains photon modes at these THz frequencies. Accordingly to the details of the chosen experimental geometry, these could be modes belonging to the same photonic branch that creates the intersubband polaritons in \Eq{pc}, but at much smaller in-plane wave vectors, or they could belong to a completely different photonic branch of the microcavity. 
For definiteness will consider the case in which the THz mode is the TEM mode of a metallic microcavity \cite{Todorov10}
and we will use different symbols for the THz modes,  calling $\alpha^{\dagger}_{\mathbf q} $ the creation operator of the low energy THz microcavity photon and $\hbar \omega_{\text{THz}}(q)$ its energy. 

Here we consider the process, depicted in Fig. \ref{Sketch}, where a polariton is first excited into the upper branch with a pump beam, and then it scatters to the lower branch by emitting a THz photon into the microcavity.
Such a coupling to the low energy THz modes will be perturbatively described by the interaction term \cite{Quattropani}:
\begin{eqnarray}
\label{Hint}
H_{\text{int}}&=&ez\sum_{ \mathbf{q}}\sqrt{\frac{\hbar\omega_{\text{THz}}(q)}{2\epsilon_0\epsilon_r SL_{\text{cav}}}}
(\alpha_{\mathbf q}^{\dagger} e^{i\mathbf{\boldsymbol{{\rho}}\, q}}+ \alpha_{\mathbf q} e^{-i\mathbf{\boldsymbol{{\rho}}\, q}}),
\end{eqnarray}
where $L_{\text{cav}}$ is the microcavity thickness.
In second quantization, $ze^{i\mathbf{\boldsymbol{{\rho}}\, q}}$ reads
\begin{eqnarray}
\label{rsc}
ze^{i\mathbf{\boldsymbol{{\rho}}\, q}}&=&\sum_{\mu,\mu'\in \{1,2\},\mathbf{k}}{z_{\mu\mu'}} \cc{\mu}{k}\ca{\mu'}{k+q},
\end{eqnarray}
where the $z_{\mu\mu'}$ are defined in \Eq{zdef}.
The terms in \Eq{rsc} with $\mu\neq\mu'$ do not couple with the THz mode. They instead give rise to the intersubband dipole, and as such they have already been considered in \Eq{Hlm}. The terms with $\mu=\mu'$ instead, usually neglected because in the weak coupling regime they give rise only to non-resonant transitions between electrons in the same subband, are here responsible for the coupling with the THz photonic mode.
Our approach of diagonalizing the Hamiltonian in \Eq{HB} and then study perturbative transitions induced between its eigenstates by the Hamiltonian in \Eq{Hint} is thus justified and does not lead to any double-counting.

The matrix element describing the process illustrated in Fig. \ref{Sketch}, in which a polariton jumps from the upper to the lower branch by emitting a THz photon, can be calculated using the interaction Hamiltonian in \Eq{Hint} between a state with an upper polariton and a state with a lower polariton plus a THz photon (using Eqs. \ref{bc} and \ref{pc}). 
We obtain
\begin{eqnarray}
\label{matel}
\langle G \lvert \alpha_{\mathbf{p}} \, p_{\text{LP}, {\mathbf q'}} H_{\text{int}}\,p^{\dagger}_{\text{UP},\mathbf{q}}\lvert G \rangle&=&{y}_{\text{UP},q}{y}_{\text{LP},q'}e\Delta z \sqrt{\frac{\hbar\omega_{\text{THz}}(p)}{2\epsilon_0\epsilon_r SL_{\text{cav}}}}
\nonumber \\&& \times\delta(\mathbf{q-q'-p}),
\end{eqnarray}
where $\lvert G \rangle=\prod_{k<k_F}\cc{1}{k}\lvert 0 \rangle$ is the Fermi ground state of the electronic system, $\lvert 0 \rangle$ is the vacuum state
and 
$
\Delta z=z_{22}-z_{11},
$
representing the distance between the mean electron positions in the two electronic subbands, is the interbranch dipole, coming from the $\mu=\mu'$ terms in \Eq{rsc}. 
As we have anticipated,  this scattering process occurs only in asymmetric systems, where $\Delta z \neq 0$.
The presence of the two Hopfield coefficients multiplying the interbranch dipole in \Eq{matel} shows how  the proposed scattering mechanism can be interpreted as a two-photon process in which one of the two photonic modes is strongly coupled.

It is apparent from Fig. \ref{Sketch} that, for a given initial upper polariton mode, there is a continuum of final modes
conserving both energy and momentum, given by the intersection between the lightcone of the THz mode irradiating from the 
initial upper polariton mode and the lower polariton branch. 
 Above threshold one of these modes will eventually win  a mode selection competition in the THz cavity and start to lase.  
The mode that wins will depend upon a number of parameters, including its scattering matrix element in  \Eq{matel}, its quality factor and its geometry, as modes propagating at different angles will incur different losses.
As the aim of the present paper is more to demonstrate the potential of a new lasing mechanism than to design a specific experiment, we will limit ourselves to consider here the forward scattering
process, shown in Fig. \ref{Sketch}, in which $\mathbf{q}$ and $\mathbf{q'}$ are parallel. Such a channel normally maximizes the scatterimg matrix element in 
\Eq{matel}, as the final mode with the larger value of $q'$ is the one with the highest matter fraction $y_{\text{LP},q'}$ and it is generally the one most likely to win the mode selection competition.
  
Now we are concerned only with three modes; the upper polariton, the lower polariton and the THz photon. Their in-plane wave vectors are fully determined by the argument above, so we can simplify the notation omitting the wave vectors for the quantities relative to these modes.

The spontaneous emission matrix element in \Eq{matel}  is not enough on its own to let us calculate the THz emission in the stimulated regime; instead we need the full many body matrix element describing a transition from a state with $N_{\text{UP}}$ upper polaritons, $N_{\text{LP}}$ lower polaritons and $N_{\text{THz}}$ THz photons to a state in which an upper polariton has become a lower one, emitting a THz photon. 
Were intersubband excitations elementary bosons, the transition rate would be exactly equal to the spontaneous rate times the bosonic enhancement rate ${N_{\text{UP}}(N_{\text{LP}}+1)(N_{\text{THz}}+1)}$.
 However, intersubband excitations are not pure bosons but composite ones \cite{Combescot07}, and the enhancement rate could thus deviate from the bosonic one at higher
excitation densities, due to the underlying fermionic degrees of freedom. In order to solve this problem, we numerically calculated the many body matrix elements using the fermionic expression of the intersubband operator given in \Eq{bc}. We do not report here all the cumbersome calculations of these coefficients, but the interested reader can find the exact calculations detailed in Ref. [\onlinecite{DeLiberato09b}]. 
It turns out that the coboson nature of the intersubband excitations modifies the bosonic enhancement by a nonbosonicity factor $B_{N_{\text{UP}}}^{N_{\text{LP}}}<1$, 
that can be calculated, for arbitrary $N_{\text{UP}}$ and $N_{\text{LP}}$, by numerically iterating a recursion relation over fermionic matrix elements. 





Assuming, for simplicity, that the lower polariton has a Lorentzian lineshape, and defining
$\Gamma_{\text{UP}}$, $\Gamma_{\text{LP}}$ and $\Gamma_{\text{THz}}$ to be the linewidths of the different modes,
the spontaneous scattering rate $\Xi_s$  for the considered interbranch dipole transition can be approximated via the Fermi golden rule  as
\begin{eqnarray}
\label{Gammasingle}
\Xi_s&=&
\frac{\omega_{\text{THz}}( \Delta ze{y}_{\text{UP}}{y}_{\text{LP}})^2 }{2\hbar\epsilon_0\epsilon_r SL_{\text{cav}}}
  \int d\omega
\frac{\Gamma_{\text{LP}}\delta(\omega_{\text{UP}}-\omega_{\text{THz}}-\omega )}{
(\omega-\omega_{\text{LP}})^2+(\frac{\Gamma_{\text{LP}}}{2})^2},\quad\,
\end{eqnarray}
where we have assumed, as is usually the case experimentally, that the 
width of the lower polariton mode is much larger than that of the THz photon mode ($\Gamma_{\text{LP}}\gg\Gamma_{\text{THz}}$). 
For the resonant, stimulated process, this becomes
\begin{eqnarray}
\label{Gammares}
\Xi(N_{\text{UP}},N_{\text{LP}},N_{\text{THz}})=\frac{ 2B_{N_{\text{UP}}}^{N_{\text{LP}}}\omega_{\text{THz}}(\Delta ze {y}_{\text{UP}}{y}_{\text{LP}})^2 }{\hbar\epsilon_0\epsilon_r SL_{\text{cav}}\Gamma_{\text{LP}}}&&\\ \times
\lbrack N_{\text{UP}}(N_{\text{LP}}+1)(N_{\text{THz}}+1)-(N_{\text{UP}}+1)N_{\text{LP}}N_{\text{THz}} \rbrack. \nonumber&&
\end{eqnarray}
If we denote the pumping rate into the upper polariton mode as $P$, we can thus write the following rate equation for the populations in the three modes
\begin{eqnarray}
\label{rateeq}
\dot{N}_{\text{UP}}&=&-\Gamma_{\text{UP}} N_{\text{UP}} -\Xi_{}(N_{\text{UP}},N_{\text{LP}},N_{\text{THz}}) +P \nonumber\\
\dot{N}_{\text{LP}}&=&-\Gamma_{\text{LP}} N_{\text{LP}}+\Xi_{}(N_{\text{UP}},N_{\text{LP}},N_{\text{THz}}) \\ 
\dot{N}_{\text{THz}}&=&-\Gamma_{\text{THz}} N_{\text{THz}} +\Xi_{}(N_{\text{UP}},N_{\text{LP}},N_{\text{THz}}).\nonumber
\end{eqnarray}

In the  dilute regime limit $B_{N_{\text{UP}}}^{N_{\text{LP}}}\simeq 1$, we can solve  \Eq{rateeq} for its steady state and compute the quantum efficiency as
\begin{eqnarray}
\label{eta_an}
\eta=\frac{{x}_{\text{UP}}^2\Gamma_{\text{THz}} N_{\text{THz}}}{P}=
\frac{{x}_{\text{UP}}^2\max\lbrack  1-\frac{\Gamma_{\text{UP}}\Gamma_{\text{THz}}}{\Xi_s P},0  \rbrack}{1+\frac{\Gamma_{\text{UP}}}{\Gamma_{\text{LP}}}},
\end{eqnarray}
where the Hopfield coefficient ${x}_{\text{UP}}$ accounts for the fact that the pump beam couples only to the photon part of the upper polariton
and we assume, as is usually the case \cite{Carosella12}, that the cavity losses exceed by far free carrier absorption losses.  
We can now write the condition for stimulated THz emission to be possible as
$\Xi(N_{\text{UP}},0,0)>\Gamma_{\text{THz}}$. 
From \Eq{Gammares}  the threshold density for stimulated emission, neglecting $B_{N_{\text{UP}}}^{0}$, is then
\begin{eqnarray}
\label{threshold}
\frac{N_{\text{UP}}}{S}&=&\frac{\Gamma_{\text{THz}}}{\omega_{\text{THz}}} \frac{\hbar\epsilon_0\epsilon_r L_{\text{cav}}\Gamma_{\text{LP}}}{2(\Delta ze {y}_{\text{UP}}{y}_{\text{LP}})^2 }.
\end{eqnarray}

\psfrag{X2}[c][b][1]{Pump power density ($\times 10^9$ W/m$^2$)}
\psfrag{Y2R}[B][B][1]{Excitation fraction}
\psfrag{Y2L}[B][B][1]{Quantum efficiency}
\psfrag{L}[B][B][1]{$L_{\text{QW}}$}
\psfrag{V}[B][B][1]{$V$}
\psfrag{D}[B][B][1]{$D$}

\begin{figure}[h!]
\begin{center}
\includegraphics[width=8.0cm]{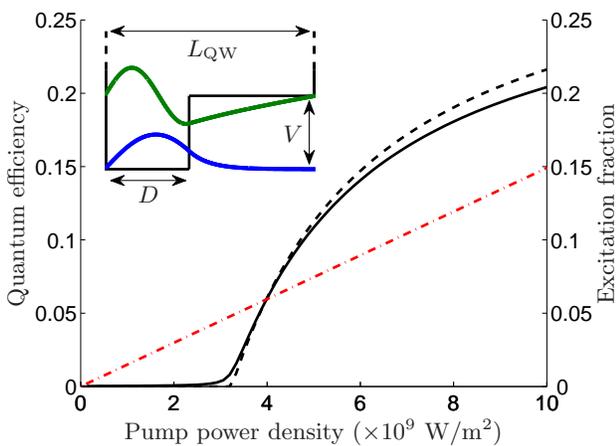}
\caption{\label{Efficiency} Black solid line: quantum efficiency  $\eta$ as a function of the applied pump power density calculated solving numerically \Eq{rateeq}, including thus nonbosonicity effects. Black dashed line: same quantity obtained using \Eq{eta_an}. Red dash-dotted line: total excitation fraction $\frac{N_{\text{UP}}+N_{\text{LP}}}{N}$. 
 Inset: wavefunctions for the first two conduction subbands for an asymmetric GaAs quantum well structure. 
Due to the quantum well asymmetry,  $\Delta z = \langle\psi_2\vert z \vert \psi_2 \rangle - \langle\psi_1\vert z \vert \psi_1 \rangle \neq 0 $. 
All the parameters are given in the text.
}
\end{center}
\end{figure}

Using this formalism to predict quantitative conversion efficiencies requires us to consider a specific example structure. In general, a compromise is needed to get a large enough $z_{12}$ to achieve the strong-coupling required to create a polariton frequency splitting in the THz range and a large enough $\Delta z$ to have an interbranch transition dipole. 
As an example we consider the simple asymmetric stepped GaAs quantum well of Fig. \ref{Efficiency} with infinite barriers. It is characterized by only three parameters, the overall width, $L_{\text{QW}}$, the height, $V$, and the position, $D$,  of the potential step. The wavefunctions shown in Fig. \ref{Efficiency} correspond to $L_{\text{QW}}=25$ nm, $D=0.4L_{\text{QW}}$ and $V=138$ meV; this gives an ISBT energy  $\hbar\omega_{12}\simeq 100$ meV and $\Delta z\simeq z_{12}\simeq 0.1L_{\text{QW}}$.  These values are not optimal, but similar ones are obtainable in a fairly large sector of the parameter space and we thus expect them to be easily achievable also in more realistic geometries.
We model a structure composed of $n_{QW} = 40$ such GaAs quantum wells each doped at an electron density $N_{2DEG} = 10^{12}$ cm$^{-2}$, that gives a resonant vacuum Rabi frequency $\Omega(q_{\text{res}})\simeq 0.1\omega_{12}$, where $q_{\text{res}}$ is the resonant wave vector such that $\omega_{\text{ph}}(q_{\text{res}})=\omega_{12}$.
Assuming  ${y}_{\text{UP}}\simeq0.6$ and ${y}_{\text{LP}}\simeq0.9$ (the values of the process represented in Fig. \ref{Sketch}), a quality factor for the THz mode \cite{Andronico08}  $\frac{\omega_{\text{THz}}}{\Gamma_{\text{THz}}}=100$ and polaritonic linewidths $\Gamma_{\text{UP}}=\Gamma_{\text{LP}}=4$ meV, appropriate for the considered doping, as described in Ref.  [\onlinecite{Luin01}], \Eq{threshold} gives $\frac{N_{\text{UP}}}{N}\simeq 0.05$, that is to say that the onset of stimulated THz emission occurs when only $\simeq 5\%$ of the electrons are in the excited subband. 
Such a low excitation density justifies, {\it a posteriori}, the approximation neglecting the nonbosonicity factor in \Eq{threshold}.

\psfrag{X3}[c][B][1]{$\omega_{\text{UP}}$ (In units of $\omega_{12}$)}
\psfrag{Y3}[B][B][1]{$\omega_{\text{THz}}$ (In units of $\omega_{12}$)}
\begin{figure}[h!]
\begin{center}
\includegraphics[width=8.0cm]{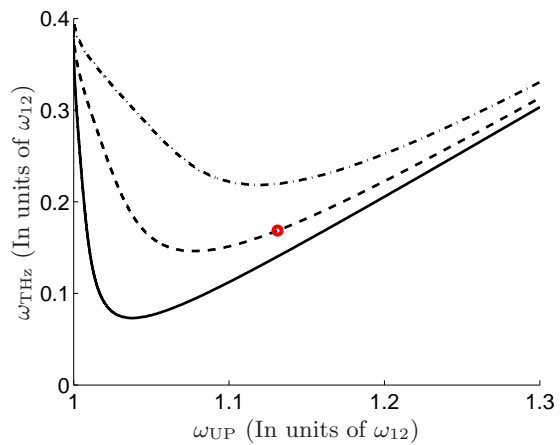}
\caption{\label{Others} Frequency of the emitted THz radiation $\omega_{\text{THz}}$, considering forward scattering, as a function of the pump frequency $\omega_{\text{UP}}$.  The different lines correspond to $\frac{\Omega}{\omega_{12}}=0.05$ (solid line), $0.075$ (dashed line) and $0.1$ (dash-dotted line).  The red circle marks the parameters considered for the numerical application in the text.}
\end{center}
\end{figure}

In Fig. \ref{Efficiency} we see the quantum efficiency as a function of the pump power obtained by numerically solving \Eq{rateeq} (solid black line) and using \Eq{eta_an}, that is, neglecting nonbosonicity effects (black dashed line) \cite{Frogley06,Gambari10}. 
In the same figure we also plot the total excitation density $\frac{N_{\text{UP}}+N_{\text{LP}}}{N}$ as a function of the pump power (red dash-dotted line).
We see that, with experimentally achievable pump powers \cite{Dynes05},  a  very high quantum conversion efficiency  is possible. 

In Fig. \ref{Others} we explore the versatility of this system by studying the way the forward-scattered emitted THz photon frequency $\omega_{\text{THz}}$ depends on the pump frequency $\omega_{\text{UP}}$, for different values of the light-matter coupling $\Omega(q_{\text{res}})$.  At lower values of $\Omega(q_{\text{res}})$ the mechanisn gives a frequency down-conversion up to a factor 
 $10$, while staying in the strong coupling regime and maintaining large conversion efficiencies. Since $\Omega(q_{\text{res}})$ can be modulated by changing the electron gas density \cite{Anappara05,Gunter09}, this offers a way of realizing widely tunable THz emitters. 

In conclusion we have shown how, through the use of asymmetric quantum wells in intersubband polariton systems, it is possible to obtain  stimulated THz emission characterized both by an extremely large quantum efficiency and a remarkable frequency tunability. 

We would like to thank I. Carusotto, L. Nguyen-th\^e, C. Sirtori and Y. Todorov for useful discussions.
C.C. is member of Institut Universitaire de France. We acknowledge support from the ANR project THINQE-PINQE.

\end{document}